\documentclass[traditabstract]{aa} 

\usepackage{graphicx}
\usepackage{txfonts}
\usepackage{color}
\usepackage{natbib}
\usepackage{verbatim}

\newcommand{\cv}{C$_{24}$\,}
\newcommand{\cs}{C$_{60}$\,}
\newcommand{\csp}{C$_{60}^+$}

\begin{document}
\title{Interstellar C$_{60}^+$}
\author{O. Bern\'e \inst{1,2}, G. Mulas \inst{3} and C. Joblin \inst{1,2}}
\institute{Universit\'e de Toulouse; UPS-OMP; IRAP;  Toulouse, France
\and CNRS; IRAP; 9 Av. colonel Roche, BP 44346, F-31028 Toulouse cedex 4, France
\and Istituto Nazionale di Astrofisica -- Osservatorio Astronomico di Cagliari -- strada 54, localitˆ Poggio dei Pini, 09012-- Capoterra (CA), Italy}

\titlerunning{Interstellar \csp}
\authorrunning{Bern\'e, Mulas, Joblin}
\date{Received August ??, 2012; accepted ??, 2012}
\abstract{Buckminsterfullerene (\cs) has recently been detected  through its infrared emission bands in the interstellar medium (ISM),
 including in the proximity of massive stars, where physical conditions could favor  the formation of the cationic form,  \csp.
 In addition, \csp was proposed as the carrier of two diffuse interstellar bands in the near-IR, although a firm identification still awaits
gas-phase spectroscopic data.
We examined in detail the \emph{Spitzer} IRS spectra of the NGC 7023 reflection nebula, at a position close (7.5'') to the illuminating B star HD 200775, and found four 
previously unreported bands at 6.4, 7.1, 8.2, and 10.5 $\mu$m, in addition to the classical bands attributed to
polycylic aromatic hydrocarbons (PAHs) and neutral \cs. These 4 bands are observed only in this region
of the nebula, while \cs\, emission is still present slightly farther away from the star, and PAH emission even farther away.
Based on this observation, on theoretical calculations we perform, and on laboratory studies, we attribute these bands to \csp.
The detection of  \csp\ confirms the idea that large carbon molecules exist in the gas phase in these environments.
In addition,  the relative variation in the  \cs\, and \csp\, band intensities constitutes a potentially powerful probe of the physical conditions in 
highly UV-irradiated regions.}

\maketitle



\section{Introduction} \label{int}

The mid-infrared (mid-IR) spectrum of galactic and extragalactic objects exhibits band emission (strongest at 3.3, 6.2, 7.7, 8.6, and 11.2 $\mu$m)
attributed to carbonaceous macromolecules, i.e., polycyclic aromatic hydrocarbons (PAHs, see recent state of the art in \citealt{job11}). In addition to PAH bands, 
{ IR emission bands} at 7.0, 8.5, 17.4, and 19.0 $\mu$m have recently been reported \citep{cam10,sel10}, and found to 
{ match quite closely the IR active bands} of buckminsterfullerene (C$_{60}$,~\citealt{kro85}), a cage-like carbon molecule. 
Carbonaceous macromolecules, {  including  PAHs}, carbon clusters, or fullerenes, are believed to play a fundamental role in the physics and chemistry
of the interstellar medium ({ISM)}, and their infrared signatures are commonly used as a tracer of physical conditions. Nevertheless, \cs is the only molecule belonging
 { to this family}, which has been specifically identified in the {  ISM}. 
In the NGC 7023 reflection nebula, \citet{sel10} have shown that \cs\, is predominantly found in the regions closest to the star. In that part of the nebula,
{ UV} irradiation is high (above $10^4$ times the interstellar standard radiation field), and PAH molecules are ionized \citep{rap05, ber07, pil12},
{ if not destroyed \citep{ber12, mon12}. One could therefore expect \csp\, to be present in these regions.}

\citet{foi94} reported evidence of interstellar \csp\, based on the detection of { two} diffuse interstellar bands (DIBs) at 9577  and 9632 \AA,
{ however this identification is still questioned considering that no spectrum of  \csp\, could be recorded yet, in conditions appropriate for DIB
identification, i.e., in gas phase and at low temperature.}
{ The IR spectrum of \csp\, was measured in a rare-gas matrix by \cite{ful93} and was found to exhibit two bands at 7.1 and 7.5 $\mu$m.
 \cite{ker12} have performed new spectroscopic measurements and suggests that the latter band is due to C$_{60}^-$, whereas the authors
 attribute a band at 6.4 $\mu$m to \csp.} \citet{mou99} derived upper limits on the abundance of \csp\, in 
NGC 7023 based on not detecting the 7.1 (and possibly wrongly attributed 7.5 $\mu$m) bands. So far, there has been no observational
evidence of any 7.1 or 6.4 $\mu$m bands in astronomical sources.

Looking carefully at the \emph{Spitzer} data of NGC 7023, we found four emission bands, 
at 6.4, 7.1, 8.2, and 10.5 $\mu$m, which are only present in the regions closest to the star. 
This also corresponds to a region where  \cs\, emission is strong. A natural carrier to explain 
these bands is \csp, and this assertion is { supported} by spectroscopic arguments that we 
discuss hereafter.

\begin{figure*}
\begin{center}
\includegraphics[width=\hsize]{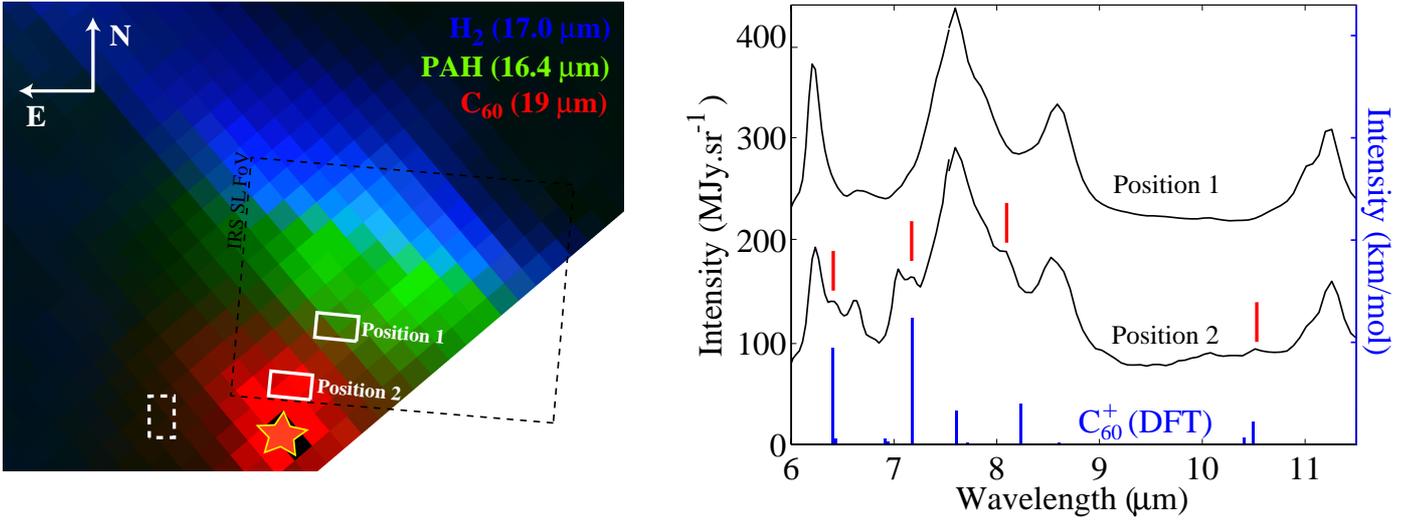}
\caption{ \emph{Left:} False-color image of the NGC 7023 nebula such as the one presented in \citet{sel10}, obtained from integrating
different components in the \emph{Spitzer}-IRS  LL spectral cube. Red is the emission integrated in the \cs\, 19$\mu$m band.
Green is the emission of the PAH 16.4 $\mu$m band. Blue is the emission integrated in the H$_2$ (0-0) S(0) 17.0 $\mu$m 
band. The white rectangles indicate the regions where the IRS-SL spectra shown in the right panel have been extracted.
{ The black dashed rectangle depicts the IRS-SL field of view.} The white dashed rectangle shows the region where \cite{sel10} 
extracted their spectrum. \emph{Right:} Spectra at positions 1 and 2
in the image. The spectrum at position 1 has been shifted up and scaled down to allow easier comparison to the spectrum at position 2.
{ Error bars are not shown here but are comparable to the width of the line (see Fig. 2)}. The red lines indicate the four newly detected bands 
attributed to \csp. The DFT calculated spectrum (with scaled wavelength positions) is shown as a bar graph in blue.
\label{Spectra}}
\end{center}
\end{figure*}

\section{Observations} \label{obs}

NGC 7023 was observed with the short-low (SL)  and long-low (LL) modules of the InfraRed Spectrograph 
(IRS, \citealt{hou04} onboard {\it Spitzer} \citep{wer04} in spectral mapping mode. { The spectral resolution of
IRS is $\lambda/\Delta \lambda=60-130$. The slit width (comparable to angular resolution) is 3.6'' for the SL 
and 10.5'' for the LL modules}. Data reduction was 
performed with the CUBISM software \citep{smi07b} and consisted in cube assembling, calibration, 
flux correction for extended sources and bad pixel removal. From the LL data cube we extracted maps
integrated in the H$_2$ line at 17.0 $\mu$m, the PAH band at 16.4 $\mu$m and the \cs\, band at 19.0 $\mu$m.
The maps are presented in Fig. 1 in a fashion similar to \citet{sel10}. From the SL data cube, we extracted two significant spectra corresponding to 
different regions in the nebula (Fig.~1). The extraction regions were 3x5 pixels (1.6''/pixel) { to improve the signal-to-noise ratio}. Position 1 corresponds to the 
cavity of atomic gas between the star and the molecular cloud at an angular distance of 21'' ($\sim$ 0.04 pc) 
from the star HD 200775 ($\alpha$=21:01:34.8, $\delta$=+68:10:07.3). { Position 2 ($\alpha$=21:01:37.3, $\delta$=+68:09:55.3) is the closest position
to the star in the IRS spectral cube  (7.5'', $\sim$ 0.015 pc from HD 200775, see Fig. 1)}. These two positions correspond 
to regions where \cs\, emission is strong in the 19.0~$\mu$m band (Fig. 1).
{ For both positions the signal-to-noise ratio is very good and ranges between 100 and 200.}

\section{Observational results} \label{obs_res}

The spectra (Fig. 1) have in common that they are dominated by bands at 6.2, 7.7, 8.6, and 11.2 $\mu$m, which are attributed to vibrational modes
of PAH molecules \citep{tie05}. In addition to these bands, the spectrum at 
position 2 shows several bands that are absent in the rest of the nebula. 
These are at $\sim$ 6.4, 6.6, 7.0, 7.1, 8.2, and 10.5 $\mu$m (Fig.~\ref{Spectra}).
{ These bands are seen in several pixels of the IRS cube and well above the 
instrumental errors (Fig. 2)}. To derive the precise parameters for these bands we fit them using Gaussian profiles  (Fig. 1) and splines to subtract
the underlying emission due to the wings of PAH bands. The positions, widths, and intensities of the bands are given in Table 1.
The 7.0 $\mu$m band has been attributed to \cs\, \citep{sel10}. The 6.6 $\mu$m band has recently been attributed to (possible) planar \cv \citep{gar11}.
 The 6.4, 7.1, 8.2, and 10.5 $\mu$m bands have not been observed or discussed yet. 
{ Since HD 200775 is a B star, only low ionization potential atoms should emit in fine structure lines (e.g. [CII], [SIII] etc.), and these
species do not have lines in this spectral range, so we exclude contamination by fine structure lines.}
These { four new bands seem spatially correlated, i.e. all of them only appear
in the regions closest to the star, which suggests a common carrier.} As shown by
 \citet{sel10}, \cs\, is also found only close to the star. { Still, the four new bands only appear in the regions that are
 closest to the star, while Fig. 1 demonstrates that \cs\, emission is more extended. This suggests that the four new bands
are carried by a species that is a product of the photoprocessing of \cs, an obvious carrier being \csp.}
In the following section we provide the spectroscopic arguments that support this observational evidence.

\section{Spectroscopy of \csp}\label{spec}

{ The only IR spectroscopic data of  \csp\, has been obtained in rare gas matrices by \citet{ful93} and
more recently by \citet{ker12}. Two bands at 7.1 and  6.4\,$\mu$m seem definitively attributed
to \csp, based on these experimental studies.}

Theory is another approach to obtaining spectroscopic data, and density functional 
theory (DFT) in particular has been shown to be effective and accurate for calculations 
on neutral \cs\, \citep{chase1992,fabian1996,iglesias-groth2011}.
However, an additional theoretical problem is that upon ionization C$_{60}$ is known to undergo 
Jahn\textendash Teller (JT) distortion \citep{cha97, ber06}. Neutral C$_{60}$ has $I_h$ symmetry, 
and is a closed\textendash shell system, and its ground electronic state is totally symmetric and nondegenerate. 
{
Upon ionization, the ground electronic state of \csp, in the adiabatic approximation, is 
five-fold degenerate and has $h_u$ symmetry. These electronic states, degenerate
at the symmetric configuration, split when the symmetry is broken, and the Jahn\textendash Teller theorem
predicts that there must be lower energy extrema, some of which must be minima, at distorted geometries.
The closeness of these electronic states means that the adiabatic approximation may break down in this
situation. Nevertheless, standard harmonic vibrational spectra can be computed in the adiabatic 
approximation around the new minima, and they will still be approximately correct if IR\textendash active
modes have a negligible component along the directions involved in the Jahn\textendash Teller effect.
This is the case for \csp, as discussed in Appendix~\ref{jtappendix}, where the reader can also find a more detailed 
description of the Jahn \textendash Teller effect and how it affects our calculations.

We performed our DFT calculations using the hybrid B3LYP exchange\textendash correlation functional 
and the 4\textendash 31g Gaussian basis set. This combination is known to yield reasonably accurate 
vibrational frequencies, after scaling with an empirical factor $\chi$ to account for the 
overestimation of the frequencies. The relatively small basis set limits the accuracy of band 
intensities, but based on the thorough tests performed by \citet{gal11}, we expect that neglecting 
dynamical JT effects is the leading limitation on vibrational band accuracy even at this level of theory.
The distorted geometry of minimum energy was obtained by optimizing with no symmetry constraints,
and turned out to have $D_{5v}$ symmetry. This is only very slightly distorted. 
Such a small change in geometry implies a correspondingly small change in interatomic forces, 
and it explains the qualitative similarity of the computed spectra of \csp\, and C$_{60}$. Of course, many 
more modes are IR\textendash active in distorted \csp\, 
due to the lowering of symmetry and correspondingly relaxed selection rules.
The properties of the most intense IR active bands of \csp\, are given in Table 2.}

\begin{table}
\caption{Properties of fitted bands for position 2}
\label{tab}
\begin{center}
\begin{tabular}{cccccc}
\hline \hline
 Position & FWHM & Intensity & Species  \\
\hline
 $\mu$m & $\mu$m & 10$^{-7}$ W~m$^{-2}$~sr$^{-1}$ & Assignment \\
\hline
 6.24		&0.13	& 	1.02 & PAH	\\
 6.43		&0.09	& 	3.68 &\csp 	\\
 6.61		&0.13	& 	3.41 &\cv 	\\
7.02		& 0.07	& 	2.92 &\cs 	 \\
 7.13		&0.11	& 	2.82 &\csp 	\\
 8.10		&0.10	&	0.91 &\csp  \\
 8.49		&0.14	&	1.83 &\cs	\\
 8.62		&0.28	& 	0.85 & PAH	\\
10.53	&0.11	& 	0.22 &\csp	\\
\hline
\end{tabular}
\end{center}
\end{table}

\begin{table}
\caption{Computed IR\textendash active bands of $D_{5v}$ \csp. Only IR active modes with an intensity above 0.1 km.mol$^{-1}$ are shown.}
\label{tab2}
\begin{center}
\begin{tabular}{cccc}
\hline \hline
Symmetry  & Unscaled  & Scaled & Intensity \\
degeneracy & frequency & wavelength$^*$ & \\

& (cm$^{-1}$) & ($\mu$m) & (km.mol$^{-1}$) \\
\hline
E$_1$ (2) & 1607 & 6.40& 94.6 \\
A$_2$ (1) & 1599 & 6.43& 5.5 \\
A$_2$ (1) & 1488 & 6.91& 5.6 \\
E$_1$ (2) & 1482 & 6.94& 2.4 \\
E$_1$ (2) & 1434 & 7.17& 123.8 \\
E$_1$ (2) & 1353 & 7.60& 33.3 \\
E$_1$ (2) & 1333 & 7.71& 1.4 \\
A$_2$ (1) & 1250 & 8.23& 20.0 \\
E$_1$ (2) & 1249 & 8.23& 39.5 \\
E$_1$ (2) & 1196 & 8.60  & 1.4 \\
A$_2$ (1) & 989 & 10.4 & 6.6 \\
E$_1$ (2) & 981 & 10.5& 22.6\\
E$_1$ (2) & 795 & 12.9 & 2.0 \\
E$_1$ (2) & 778 & 13.2 & 6.8 \\
E$_1$ (2) & 763 & 13.5 & 0.5 \\
A$_2$ (1) & 592 & 17.4 & 3.5 \\
E$_1$ (2) & 588 & 17.5 & 0.3 \\
A$_2$ (1) & 553 & 18.6& 19.8 \\
E$_1$ (2) & 412 & 25.0& 12.1 \\
E$_1$ (2) & 364 & 28.2 & 5.0 \\
A$_2$ (1) & 361 & 28.5 & 0.6 \\
\hline
\multicolumn{4}{l}{$^*$ {\tiny using an empirical scaling factor $\chi=0.9725$}}\\
\end{tabular}
\vspace{-0.5cm}
\end{center}
\end{table}

\begin{figure}
\begin{center}
\includegraphics[width=4cm]{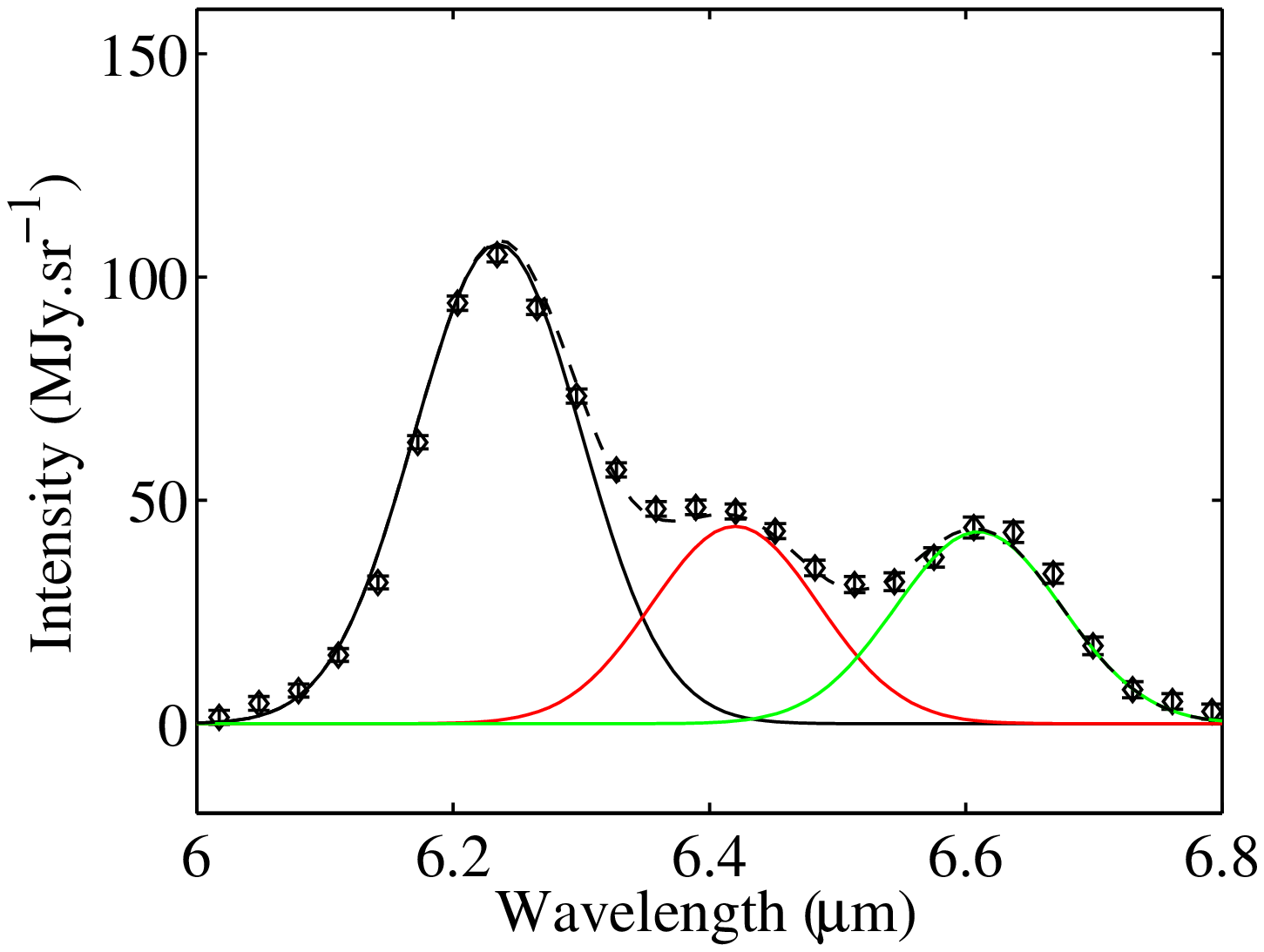}
\includegraphics[width=4cm]{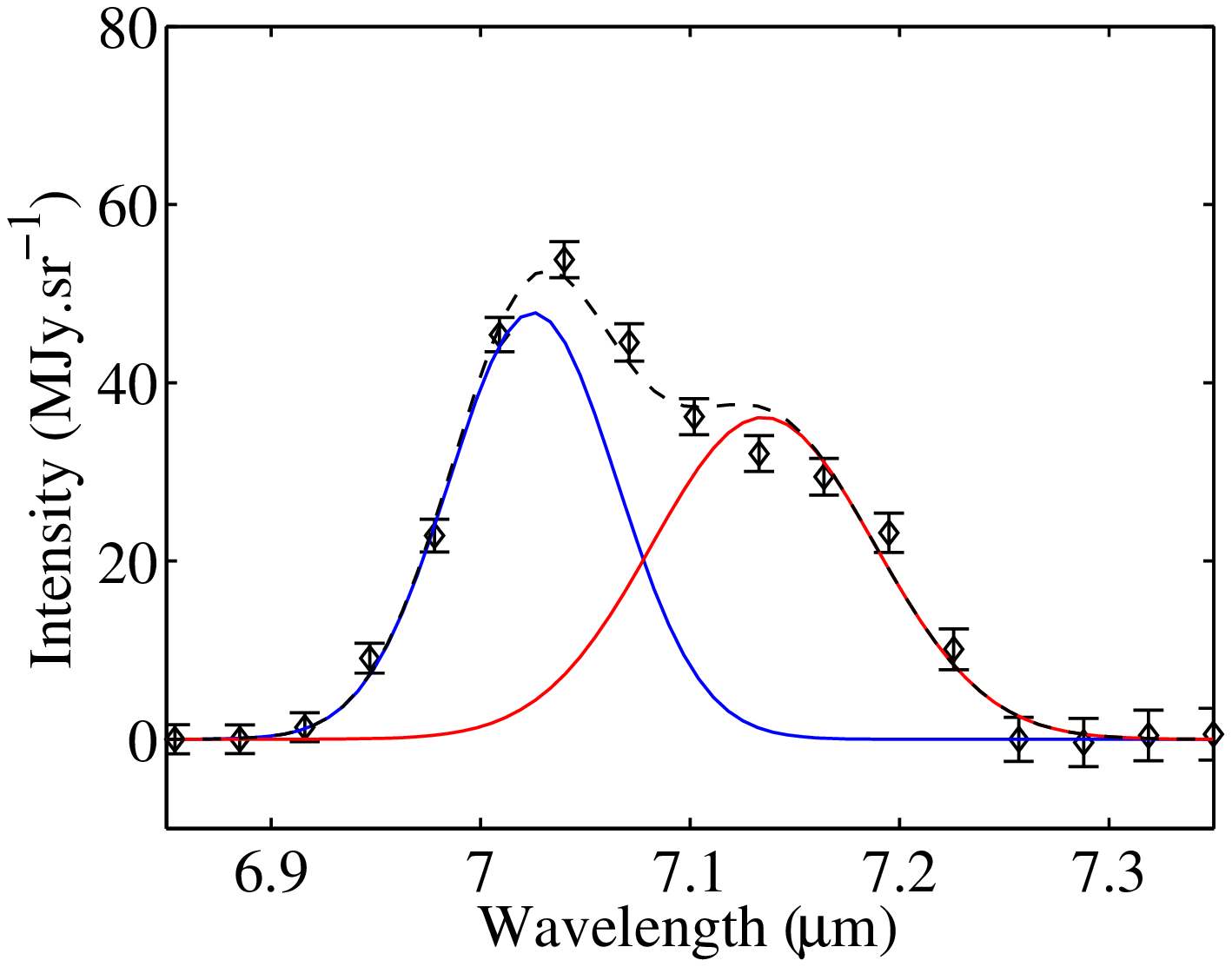}
\includegraphics[width=4cm]{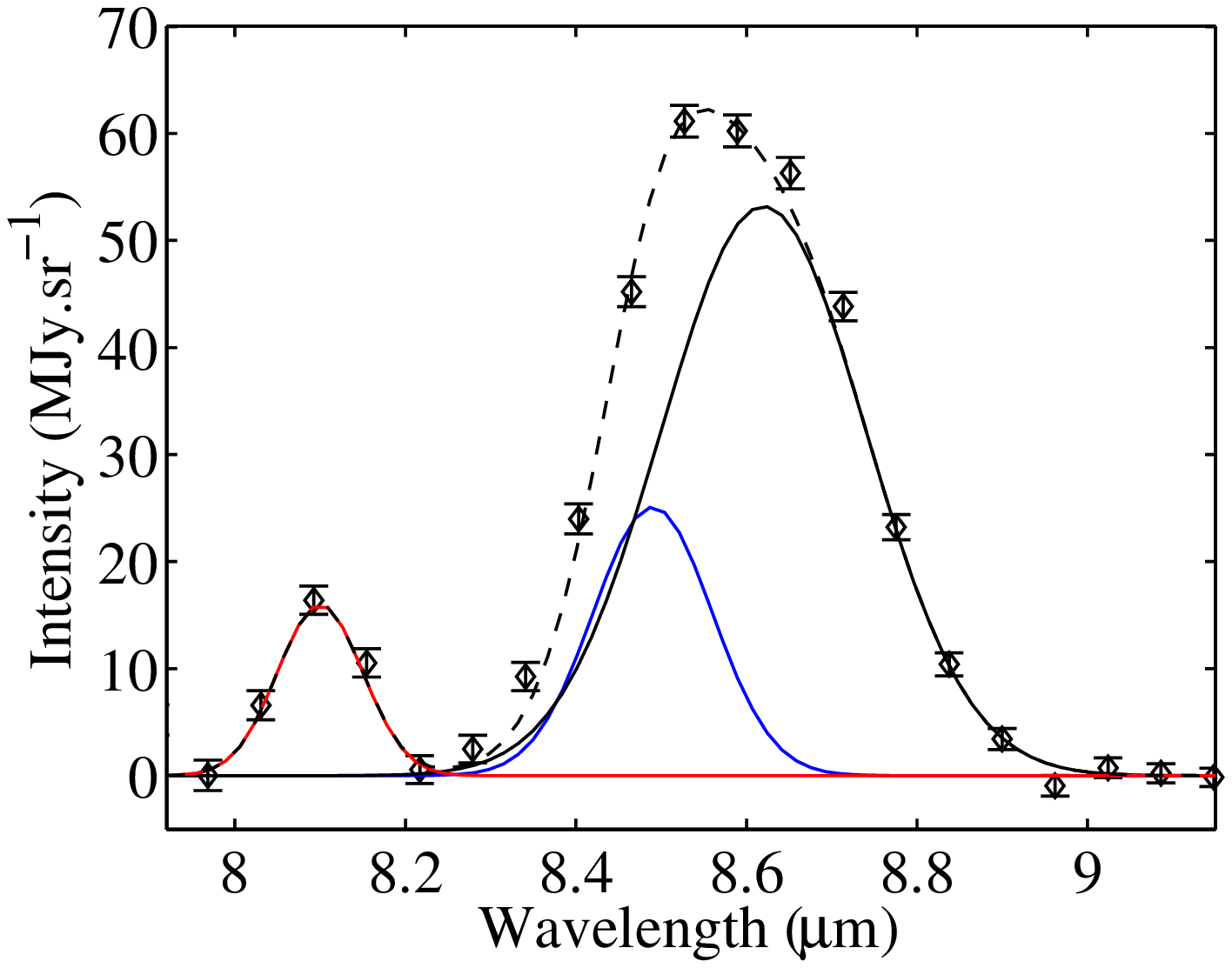}
\includegraphics[width=4cm]{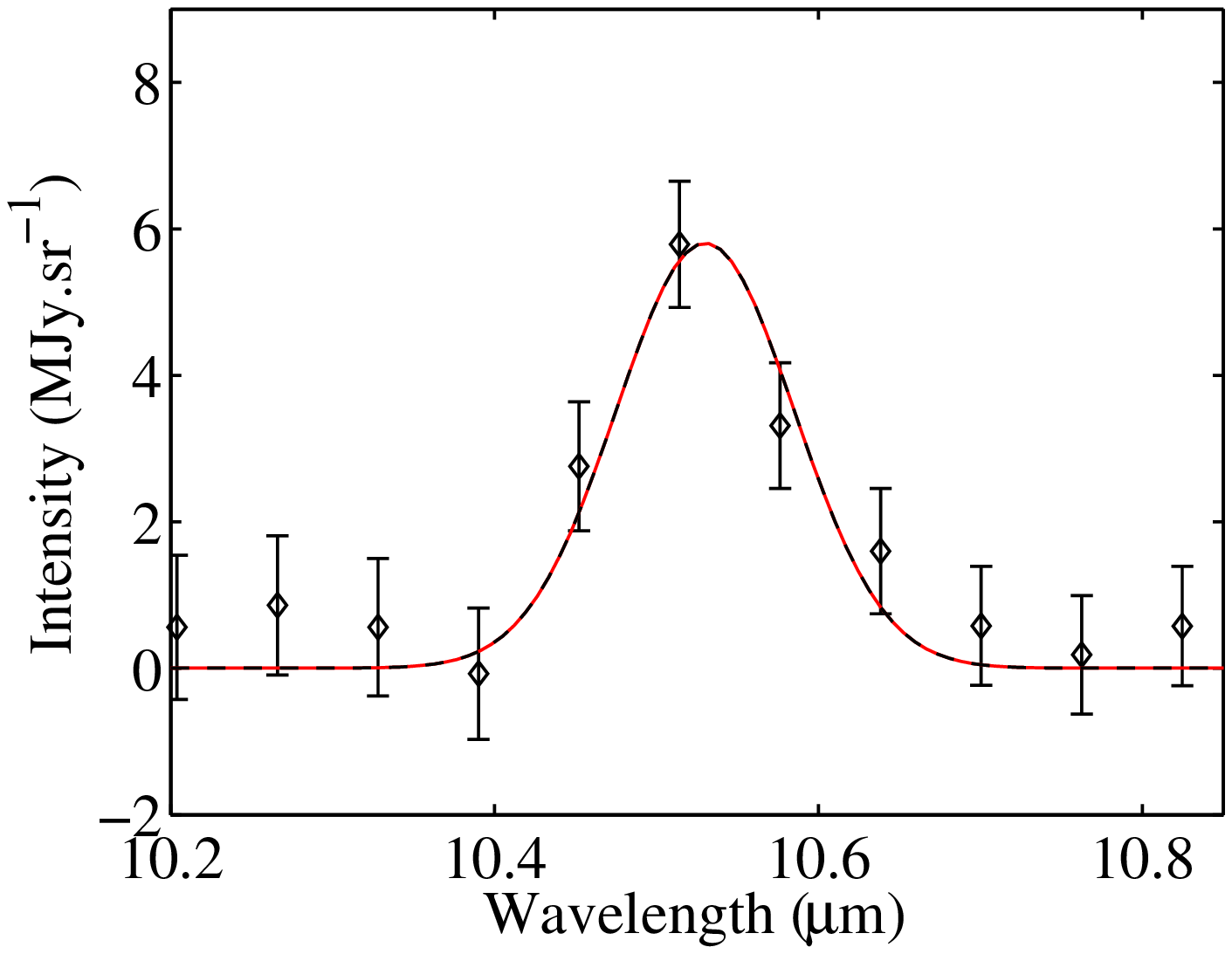}
\caption{Fits (dashed line) to the observed bands (diamonds) in position 2. The continuum has been subtracted. 
Red solid lines denote emission attributed to \csp, blue to \cs, green to \cv, and black to PAHs.
\label{fits}}
\end{center}
\end{figure}
\vspace{-0.5cm}

\section{\csp\, in NGC 7023}

\subsection{ Identification of  \csp in NGC 7023 \label{comp}}

As mentioned in the previous section, it is necessary to correct the calculated frequencies by an empirical scaling factor. 
In the case of PAHs, this is usually done by comparing the computed frequencies to the ones measured in the laboratory at 
low temperature in rare gas matrices (see \citealt{bau97} for a case-study on PAHs). Therefore we use the  IR  absorption  
spectrum  of \csp\, that was measured in Ne matrices by Kern et al. (2012) to calibrate our DFT calculations. The two bands definitively 
attributed to \csp\, by \citep{ker12} are found at 1550 and 1406 cm$^{-1}$ and correspond to the strongest bands, predicted by theory at 1607 and 
1434 cm$^{-1}$. This implies a respective scaling  factor of 0.9645 and 0.9805. We adopt an average
value $\chi=0.9725$. The corrected positions are given in Table 2 and the resulting spectrum is shown in Fig. 1. 
After scaling, the five strongest IR bands fall at wavelengths of  6.40, 7.17, 7.60, 8.23 and 10.50 $\mu$m. 
Four of these are very close (within 2\%) to the positions of the four new  bands detected in NGC 7023 (6.43, 7.13, 8.10, 10.53 $\mu$m). 
The observed match is very good considering that other factors are expected to affect the band positions, in particular band shift 
due to anharmonic coupling in hot emitting molecules \citep{job95}.
The nondetection of the {7.6} $\mu$m band in the observations is not surprising, since it is most likely 
hindered by the strong PAH emission at the same position (Fig. 1).
Based on the observational (Sect. \ref{obs_res}) and spectroscopic arguments presented in this Letter, 
we therefore argue that there is strong evidence for the presence of \csp\, in NGC 7023.
{Further discussion of the relative band intensities would require an emission model. Attempts to
build such a model for  \cs have been reported by \citet{sel10} and then by \citet{bers12}. However, the
authors could not account for the observed relative band intensities. This is likely due to uncertainties in the
intrinsic IR band strengths that were obtained by DFT calculations. The same certainly holds for \csp\, for the reasons
given in Sect.\,\ref{spec}. Still, the detection of  \csp\ { strongly} supports the idea that \cs\ is in the
gas phase \citep{sel10} and not in solid phase as initially suggested by \citet{cam10}}.

\subsection{Abundance}
Using a simple energetic consideration, we can derive an estimation of the \csp abundance.
We write that \cs\, and \csp\, relax in the IR  all the energy
they absorb in the UV, which is true if IR emission is the only relaxation channel. 
We can further assume that the UV absorption cross-section of neutral and cationic species are similar, which is the case
 for PAHs at energies higher than 10\,eV \citep{cec08}. In this case, the ionization fraction of \cs
 can be directly derived from the integrated IR flux in the \cs and \csp bands. Using the values in Table 1 we find
 $4.7\times10^{-7}$ Wm$^{-2}$sr$^{-1}$ for \cs\, and $7.7\times10^{-7}$ Wm$^{-2}$sr$^{-1}$ for \csp.
 However, these values have to be corrected by the emission at longer wavelengths. In the case
 of  \cs , we therefore add the emission in the 17.4 and 18.9\,$\mu$m bands, measured in the 
 IRS2 LL spectrum for position 2 (respectively, 2.6 and 5.2 $\times 10^{-7}$ Wm$^{-2}$sr$^{-1}$).
 This leads to a total value of
 $12.5\times10^{-7}$ Wm$^{-2}$sr$^{-1}$ for the total IR flux. In the case of \csp\,, we do not consider
 emission at longer wavelengths in line with DFT calculations, which predict much weaker bands.
{ Thus, we obtain that the ionization fraction of \cs is 38\%.}
If we consider a maximum abundance of \cs\, in NGC7023 of $1.7 \times 10^{-4}$ 
of the elemental carbon \citep{ber12}, this implies an abundance of  $\sim 1.0 \times 10^{-4}$ of the elemental carbon abundance for \csp. 
This is a factor at least 10 lower than the value derived by \cite{foi94} for the diffuse interstellar medium, assuming
that \csp is the carrier of the two DIBs  at 9577 and 9632 \AA. Since the DIB identification is still not firmly established, it is difficult
to discuss this discrepancy further.

\subsection{Taking the next step, \csp\, as a tracer of physical conditions}
\citet{sel10} report the detection of the \cs\, band at 7.0 $\mu$m in NGC 7023.
In their spectrum, the bands of \csp\, are not present. Since this spectrum was taken in 
a region farther away from the star (Fig. 1) this suggests that the \cs\, emission extends
farther away from the star than \csp\, emission. 
These relative variations can be attributed to the photochemical evolution of \cs\,
{ resulting from the competition between ionization by UV photon and recombination
with electrons. Assuming one can quantify these two processes, and also have
a proper description of the photophysics in these systems, this would allow using 
the ratio between the IR flux of \cs\, and \csp\, as a tracer of local physical conditions. There 
already exists such an approach using PAH
bands \citep{gal08}, but only empirical laws can be used since the PAH population is poorly
characterized. } Furthermore, information on the  \cs and \csp\, bands is unique for regions 
of high UV radiation fields in which other molecular tracers  may not survive.

\section{Conclusion}

After studying the mid-IR spectra of the NGC 7023 nebula, we have found spectral signatures 
at 6.4, 7.1, 8.1, and 10.5 $\mu$m, which we attribute to the cationic form of \cs\, (\csp). 
This is the largest cation know in space so far. \csp\, has been proposed as a DIB carrier,
{ and our identification supports this proposal. This is also { clear evidence
for the presence of large carbon molecules in the gas phase in the ISM.}
The detection of \csp\, in emission also opens the possibility of using the ratio between
the \cs\, and \csp\, IR bands as a tracer of physical conditions in interstellar and circumstellar
environments. This may prove useful in the framework of forthcoming 
infrared missions, such as JWST and SPICA.

\begin{acknowledgements}
We acknowledge the French National Program Physique et Chimie du Milieu Interstellaire
for its support.
\end{acknowledgements}

\bibliographystyle{aa}
\bibliography{biblio.bib}

{
\appendix
\section{Details on \csp spectroscopy} \label{jtappendix}
Neutral C$_{60}$ has $I_h$ symmetry, its highest occupied molecular orbital (HOMO)
is five\textendash fold degenerate, it has $h_u$ symmetry and is fully occupied, resulting
is a closed\textendash shell ground state that is totally symmetric and nondegenerate. 
Upon ionization, the hole in the $h_u$ HOMO yields a five\textendash fold degenerate 
$h_u$ overall electronic state.
This undergoes spontaneous symmetry-breaking due to the Jahn\textendash Teller (JT) effect 
\citep{cha97, ber06}. 
The degeneracy of the electronic state is lifted by distorting the molecule to a lower 
symmetry along some of its normal modes, which are determined by symmetry and called JT\textendash active. 
For \csp\,, JT\textendash active modes are those of 
$H_g$, $G_g$, and $A_g$ symmetry. The $A_g$ modes only shift the total energy, without reducing symmetry.
The $H_g$, $G_g$ modes instead break the $I_h$ symmetry, and produce a
multisheet adiabatic potential energy surface, with a conical intersection in the symmetric
geometry and extrema in configurations of $D_{5v}$ and $D_{3v}$ lowered symmetry.
DFT predicts that the $D_{5v}$ geometries should be the absolute minima, with the $D_{3v}$ ones being
shallow transition states \citep{sai02}. However, high-resolution photo\textendash electron spectroscopy seems to hint that 
the $D_{3v}$ geometry could be the real minimum \citep{can02}.
When JT\textendash distorted minima are deep with respect to vibrational energy (\emph{static} 
JT\textendash effect), the adiabatic approximation holds in its vicinity, and standard harmonic vibrational 
analysis is applicable. Conversely, if equivalent minima are separated by negligible potential barriers, 
the molecule can tunnel among equivalent minima, mixing the near\textendash degenerate electronic states
(\emph{dynamic} JT\textendash effect), and the adiabatic approximation is not applicable. The resulting 
vibronic states recover the full initial symmetry of the problem.
In this case,
a much more complex calculation, dropping the adiabatic approximation, is needed for accurate results.
A comparable, but much simpler case of dynamical JT effect is the cation of corannulene,
{ C$_{20}$H$_{10}^{+}$}, which can be regarded as a fragment of C$_{60}$ with peripheral bonds saturated
by H atoms. This was studied by 
\citet{gal11}, who compared experimental infrared, multiphoton dissociation (IRMPD) spectra with
a plain DFT harmonic vibrational analysis at the distorted geometry of minimum energy. This was expected
to be the worst possible comparison, since in IRMPD experiments, vibrational energy is increased until
the dissociation threshold is reached. This corresponds to energy values that are much higher than all 
barriers among equivalent minima, thereby maximizing dynamical JT effects. Despite this, 
 the experimental and theoretical spectra do qualitatively agree, allowing for accurate
band identification. Bands that are most 
significantly mispredicted, in position and intensity (but still identifiable in the laboratory spectrum),
are those whose normal modes imply displacement along JT\textendash active modes, i.~e. those that 
move the molecule from a minimum in the direction of another one, or to the conical intersection.

In the light of this, we did a similar analysis for \csp, finding the distorted geometry of minimum energy and
computing harmonic vibrational spectra there, thereby neglecting dynamical JT effects. We performed DFT 
calculations using both the Gaussian version 03.d2 and NWChem version 6.1 codes, and obtained 
very nearly identical results. Optimization led to the  $D_{5v}$ geometry. We also optimized the geometry 
of \csp with the constraint of $I_h$ symmetry, obtaining a JT stabilization energy of $\sim 70$~meV, 
which is consistent with previous calculations \citep{sai02}.
The $D_{5v}$ distorted geometry, when compared with the symmetric one, appears to be distorted almost exclusively 
along normal modes of $H_g$ symmetry, with changes in bond lengths of 
a few m\AA\ and bond angles by less than a degree (maximum).
}

\end{document}